\documentclass[12pt]{article}
\usepackage{amssymb}
\usepackage{amsmath}
\def\a{\alpha}

\def\d{\delta}

\def\la{\lambda}

\def\be{\begin{equation}}
\def\ee{\end{equation}}
\def\arr{\begin{array}{rll}}
\def\ea{\end{array}}
\def\bea{\begin{eqnarray}}
\def\eea{\end{eqnarray}}

\def\N2{$N{=}2$}

\def\>{\rangle}
\def\<{\langle}
\def\+{\dagger}
\def\={\ =\ }

\begin{document}
\renewcommand{\thefootnote}{\arabic{footnote}}
\begin{titlepage}
\setcounter{page}{0}
\begin{flushright}
LMP-TPU-- 3/14  \\
\end{flushright}
\vskip 1cm
\begin{center}
{\LARGE\bf Dynamical realizations of $\mathcal{N}=1$ }\\
\vskip 0.5cm
{\LARGE\bf $l$-conformal Galilei superalgebra}\\
\vskip 1cm
$
\textrm{\Large Ivan Masterov\ }
$
\vskip 0.7cm
{\it
Laboratory of Mathematical Physics, Tomsk Polytechnic University, \\
634050 Tomsk, Lenin Ave. 30, Russian Federation}
\vskip 0.7cm
{\it
Department of Physics, Tomsk State University, 634050 Tomsk, \\
Lenin Ave. 36, Russian Federation}
\vskip 0.7cm
{E-mail: masterov@tpu.ru}

\end{center}
\vskip 1cm
\begin{abstract} \noindent
Dynamical systems which are invariant under $\mathcal{N}=1$ supersymmetric extension of the
$l$-conformal Galilei algebra are constructed. These include a free $\mathcal{N}=1$ superparticle which is governed by higher derivative equations of motion and an $\mathcal{N}=1$ supersymmetric Pais-Uhlenbeck oscillator for a particular choice of its frequencies. A Niederer-like transformation which links the models is proposed.

\end{abstract}


\vskip 0.5cm


\end{titlepage}

\noindent
{\bf 1. Introduction}
\vskip 0.5cm

In recent years nonrelativistic (super)conformal algebras have attracted considerable attention \cite{Henkel_2}-\cite{Hosseiny}. On the one hand, this interest originates from the current investigation of the nonrelativistic version of the AdS/CFT-correspondence \cite{Son,McGreevy}. On the other hand, this research is motivated by the desire to construct new integrable models and explore novel correlations.

As is well-known, the Galilei algebra is relevant for physics in flat nonrelativistic spacetime. Conformal extension of the Galilei algebra is feasible and, moreover, it is not unique \cite{Henkel}-\cite{Negro_2}. In general, conformally extended Galilei algebra involves $(2l+1)$ vector generators where $l$ is a positive integer or half-integer. Such an extension is called the $l$-conformal Galilei algebra \cite{Negro_1}. The first two options $l=\frac{1}{2}$ and $l=1$, which are known in the literature as the Schr\"{o}dinger algebra \cite{Niederer_1} and the conformal Galilei algebra \cite{Lukierski_1,Havas}, have been the focus of most studies \cite{Henkel_2}-\cite{Galajinsky_7}, \cite{Duval_2}-\cite{Fedoruk_2}. More recently, various aspects of $l>1$ conformal Galilei symmetry have been extensively investigated. These include
the construction of dynamical realizations \cite{Gomis}-\cite{Gonera_1}, \cite{Aizawa_3}, the study of admissible central and infinite dimensional extensions \cite{Galajinsky_3,Masterov,Hosseiny}, the analysis of supersymmetric generalizations \cite{Masterov,Aizawa_1,Aizawa_2}, the investigation of irreducible representations \cite{Aizawa_4, Lu} and the possible twist deformations \cite{Dasz}.

The Galilei algebra can be obtained from the Poincar\'{e} algebra by the nonrelativistic contraction \cite{Inonu}. Likewise, the so-called Newton-Hooke algebra \cite{Bacry,Gibbons} can be derived from the (anti) de Sitter algebra. A specific feature of the Newton-Hooke algebra is that its structure relations involve the nonrelativistic cosmological constant $\Lambda=\mp\frac{1}{R^2}$, where $R$ is the characteristic time which is proportional to the radius of the parent (anti) de Sitter spacetime \cite{Gibbons}. The Newton-Hooke algebra also admits the $l$-conformal extension, which, however, is isomorphic to the $l$-conformal Galilei algebra \cite{Galajinsky_3,Negro_1,Negro_2}. The change of the basis in the $l$-conformal Galilei algebra
\bea\label{change1}
K_{-1}\rightarrow K_{-1}\pm \frac{1}{R^2}K_{1},
\eea
where $K_{-1}$ is the generator of time translations and $K_{1}$ is the generator of special conformal transformations, leads to the structure relations of the $l$-conformal Newton-Hooke algebra with negative (upper sign) or
positive (lower sign) cosmological constant. By this reason it is customary to speak about realizations of one and the same algebra in a flat spacetime and in the Newton-Hooke spacetime \cite{Gibbons}.

Supersymmetric extensions of the $l=\frac{1}{2}$-conformal Galilei algebra have been studied in \cite{Henkel_1,Galajinsky_7,Galajinsky_4,Gomis_1,Leblanc}. In particular, such supersymmetry was revealed in a nonrelativistic spin-$\frac{1}{2}$ particle, the nonrelativistic limit of the Chern-Simons matter systems, and quantum many-body mechanics. $\mathcal{N}$-extended version was systematically studied in \cite{Duval_3}. More recently, in \cite{Masterov,Aizawa_1,Aizawa_2} various supersymmetric extensions of the $l$-conformal Galilei algebra were constructed for the case of arbitrary $l$, but their dynamical realizations remain completely unexplored. The purpose of this work is to construct new dynamical realizations of $\mathcal{N}=1$ supersymmetric extension of the $l$-conformal Galilei algebra in the basis chosen in \cite{Aizawa_1}.

The paper is organized as follows. In Section 2, we recall the basic facts about the $l$-conformal Galilei algebra. In Section 3 and Section 4, we construct dynamical realizations of $\mathcal{N}=1$ $l$-conformal Galilei superalgebra in flat superspace and in Newton-Hooke superspace, respectively. In Section 5, we summarize our results and discuss possible further developments. An infinite-dimensional generalization of the $\mathcal{N}=1$ $l$-conformal Galilei algebra in $d=1$ is discussed in Appendix A.

\vskip 0.5cm
\noindent
{\bf 2. The $l$-conformal Galilei algebra}
\vskip 0.5cm

Let us recall the structure of the $l$-conformal Galilei algebra. Besides the generators $K_{-1}$ and $K_1$ mentioned above, it involves the generator of dilatations $K_{0}$, the chain of vector generators $C_i^{(n)}$ with $n=0,1,..,2l$, and the generators of spatial rotations $M_{ij}$. The structure relations read
\begin{align}\label{algebraG}
&
[K_{p},K_{m}]=(m-p)K_{p+m}, &&  [K_{p},C^{(n)}_i]=(n-l(p+1))C_i^{(p+n)},
\nonumber\\[2pt]
&
[M_{ij},C^{(n)}_k]=\delta_{ik} C^{(n)}_j-\delta_{jk} C^{(n)}_i, && [M_{ij},M_{kl}]=\delta_{ik} M_{jl}+\delta_{jl} M_{ik}-
\delta_{il} M_{jk}-\delta_{jk} M_{il}.
\end{align}

The algebra admits the central extension whose form depends on whether $l$ is even or odd \cite{Galajinsky_3}
\bea\label{MCE}
[C_i^{(n)},C_j^{(m)}]=(-1)^n n!m!\lambda_{ij}\d_{n+m,2l}M,
\eea
where
\bea
\lambda_{ij}=\left\{
\begin{aligned}
\d_{ij},&\qquad i,j=1,2,..,d,&\qquad\mbox{for half-integer $l$};\\
\epsilon_{ij},&\qquad i,j=1,2,&\qquad\mbox{for integer $l$},\\
\end{aligned}
\right.
\eea
$\epsilon_{12}=1$ and $M$ is the central charge.
In dynamical realizations the central charges correspond to physical parameters of a systems. As we shall see below in the Sections 3 and 4, the central charges play the crucial role in constructing the dynamical realizations.

\vskip 0.5cm
\noindent
{\bf 3. Dynamical realization of $\mathcal{N}=1$ $l$-conformal Galilei superalgebra}
\vskip 0.5cm

Let us consider $\mathcal{N}=1$ supersymmetric extension of the $l$-conformal Galilei algebra presented in \cite{Aizawa_1}. In addition to the generators considered in the preceding section, it involves the supersymmetry generator $G_{-\frac{1}{2}}$, the generator of superconformal transformations $G_{\frac{1}{2}}$ and the fermionic partners $L_i^{(n)}$, with $n=0,1,..,2l-1$, of the vector generators. Along with (\ref{algebraG}) the nonvanishing (anti)commutation relations of the superalgebra include
\begin{align}\label{algebraGN1}
&
\{G_r,G_s\}=2iK_{r+s}, && [K_p,L_i^{(m)}]=(m-(l-1/2)(p+1))L_i^{(p+m)},
\nonumber\\[2pt]
&
\{G_r,L_i^{(n)}\}=iC_i^{(n+r+1/2)}, && [G_r,C_i^{(n)}]=\left(n-2l\left(r+1/2\right)\right)L_i^{(r+n-1/2)},
\nonumber\\[2pt]
&
[K_p,G_r]=\left(r-\frac{p}{2}\right)G_{n+r}, && [M_{ij},L_k^{(n)}]=\d_{ik}L_j^{(n)}-\d_{jk}L_i^{(n)}.
\end{align}

The structure relations (\ref{algebraG}) and (\ref{algebraGN1}) are compatible with (\ref{MCE}) only if the anticommutators of the fermionic vector generators are modified as follows \cite{Aizawa_1}:
\bea\label{MCEN1}
\{L_i^{(n)},L_j^{(m)}\}=i(-1)^{n}n! m!\lambda_{ij}\d_{n+m,2l-1}M.
\eea

As the first step, let us check that all the scalar generators in the superalgebra, as well as space rotations, can be realized as quadratic combinations of the bosonic and fermionic vector generators. Indeed, choosing the ansatz for $K_n$
\bea
K_n=\sum_{k,m=0}^{2l}\alpha(k,m;n)\lambda_{ij}C_i^{(k)}C_j^{(m)}+\sum_{k,m=0}^{2l-1}\beta(k,m;n)\lambda_{ij}L_i^{(k)}L_j^{(m)},
\nonumber
\eea
one can unambiguously fix the constants $\alpha(k,m;n)$ and $\beta(k,m;n)$ by imposing the structure relations of the superalgebra. The explicit form of the
generators obtained in this way is
\begin{align}\label{biprod}
&
K_n=\sum_{k=0}^{2l}\frac{\alpha_k}{2} (k-l(n+1))\lambda_{ij}C_i^{(2l-k)}C_j^{(k+n)}+\sum_{k=0}^{2l-1}\frac{\beta_k}{2}(k-(l-1/2)(n+1))\lambda_{ij}L_i^{(2l-1-k)}L_j^{(k+n)},
\nonumber\\
&
G_r=\sum_{k=1}^{2l}\alpha_k k\lambda_{ij} C_i^{(2l-k+r+1/2)}L_j^{(k-1)},\quad\;
M_{ij}=\sum_{k=0}^{2l}\alpha_k C_i^{(2l-k)}C_j^{(k)}+\sum_{k=0}^{2l-1}\beta_k L_i^{(2l-k-1)}L_j^{(k)}.
\end{align}
Note  that for integer $l$ the generator of rotation reads
\bea\label{12}
M_{12}=-\sum_{k=0}^{2l}\frac{\alpha_k}{2} C_i^{(2l-k)}C_i^{(k)}-\sum_{k=0}^{2l-1}\frac{\beta_k}{2} L_i^{(2l-k-1)}L_i^{(k)},
\eea
where we denoted
\bea
\alpha_k=\frac{(-1)^{2l+k}}{Mk!(2l-k)!},\qquad\quad \beta_k=\frac{i(-1)^{2l+k-1}}{Mk!(2l-k-1)!}.
\nonumber
\eea
For $\mathcal{N}=2$ supersymmetric extensions of the $l$-conformal Galilei algebra the analogs of (\ref{biprod}) and (\ref{12}) are presented in \cite{Aizawa_1}, \cite{Aizawa_2}. Throughout the work the summation over repeated spatial indices is understood.

Thus if one succeeds in constructing a system with conserved vector charges obeying (\ref{MCE}) and (\ref{MCEN1}) with respect to some graded Poisson bracket, one can automatically produce additional integrals of motion by making use of (\ref{biprod}) and (\ref{12}). Together with $C_i^{(n)}$, $L_i^{(n)}$ they will obey the structure relations (\ref{algebraG}), (\ref{algebraGN1}) with respect to the same bracket.

Let us construct such a system by applying the method of nonlinear realizations \cite{Coleman_1,Coleman_2} to the subalgebra formed by $K_{-1}$, $C_i^{(n)}$, $L_i^{(n)}$, and $M$. To this end, one starts with a generic subgroup element $e^{aK_{-1}}e^{\zeta_i^{(n)}C_i^{(n)}}e^{i\xi_i^{(n)}L_i^{(n)}}e^{\chi M}$, where $a,\zeta_i^{(n)},\xi_i^{(n)},\chi$ are parameters, and considers the transformation of the space
\bea\label{spaceN1}
G=e^{tK_{-1}}e^{x_i^{(n)}C_i^{(n)}}e^{i\psi_i^{(n)} L_i^{(n)}}e^{\varphi M},
\eea
parametrized by the coordinates $t,x_i^{(n)},\psi_i^{(n)},\varphi$, which is generated by the left multiplication with the subgroup element. It is assumed that $\xi_i^{(n)}$ and $\psi_i^{(n)}$ anticommute with $L_i^{(n)}$ as well as with each other. The resulting infinitesimal coordinate transformations read
\bea
&&
\d t=a,\qquad\d x_i^{(n)}=\sum_{k=n}^{2l}\frac{(-1)^{k-n}k!}{n!(k-n)!}t^{k-n}\zeta_i^{(k)},\qquad\d\psi_i^{(n)}=\sum_{k=n}^{2l-1}\frac{(-1)^{k-n}k!}{n!(k-n)!}t^{k-n}\xi_i^{(k)},
\nonumber
\\[2pt]
&&\label{trN1}
\d \varphi=\chi+\sum_{n=0}^{2l}\sum_{k=n}^{2l}\frac{(-1)^{k}k!(2l-n)!}{2(k-n)!}t^{k-n}\zeta_i^{(k)}\la_{ij}x_j^{(2l-n)}+
\\[2pt]
&&
\qquad\qquad\qquad\qquad\qquad+i\sum_{n=0}^{2l-1}\sum_{k=n}^{2l-1}\frac{(-1)^{k} k!(2l-n-1)!}{2(k-n)!}t^{k-n}\xi_i^{(k)}\la_{ij}\psi_j^{(2l-n-1)}.
\nonumber
\eea

Then one constructs the Maurer-Cartan one-forms
\bea
G^{-1}dG=\omega_K K_{-1}+\omega_{i}^{(n)}C_i^{(n)}+i\tilde{\omega}_i^{(n)}L_i^{(n)}+\omega_M M,
\nonumber
\eea
where
\begin{align}\label{MC}
&
\omega_K=dt,\quad\qquad\omega_i^{(n)}=dx_i^{(n)}+(n+1)x_i^{(n+1)}dt,\quad\qquad \tilde{\omega}_i^{(n)}=d\psi_i^{(n)}+(n+1)\psi_i^{(n+1)}dt,
\\[2pt]
&
\omega_M=d\varphi+\frac{\la_{ij}}{2}\left(\sum_{n=0}^{2l}(-1)^{n}\omega_i^{(n)}x_j^{(2l-n)}n!(2l-n)!+i\sum_{n=0}^{2l-1}(-1)^n\tilde{\omega}_i^{(n)}\psi_j^{(2l-n-1)}n!(2l-n-1)!\right),
\nonumber
\end{align}
which hold invariant under all the transformations (\ref{trN1}). By definition, $x_i^{(2l+1)}=\psi_i^{(2l)}=0$.

In general, one can either reduce the number of degrees of freedom or obtain the dynamical equations of motion by setting some of the Maurer-Cartan one-forms to vanish \cite{Ivanov}. Let us choose the restrictions
\bea\label{restr}
\omega_i^{(n)}=0,\qquad \tilde{\omega}_i^{(n)}=0,
\eea
which allow us to exclude all the vector variables except for $x_i^{(0)}\equiv x_i$ and $\psi_i^{(0)}\equiv\psi_i$. Then taking $t$ to be a temporal coordinate, from (\ref{restr}) one obtains the constraints
\bea\label{constr}
x_i^{(n)}=\frac{(-1)^n}{n!}\frac{d^n x_i}{dt^n},\qquad\psi_i^{(n)}=\frac{(-1)^n}{n!}\frac{d^n\psi_i}{dt^n},
\eea
as well as the dynamical equations of motion
\bea\label{EOM}
\frac{d^{2l+1}x_i}{dt^{2l+1}}=0,\qquad\frac{d^{2l}\psi_i}{dt^{2l}}=0.
\eea

Note that the equations (\ref{EOM}) can be derived from the action functional
\bea\label{action}
S=\int (-1)^{2l+1} \omega_M=\frac{1}{2}\int dt\,\la_{ij}\left(x_i\frac{d^{2l+1}x_j}{dt^{2l+1}}-i\psi_i\frac{d^{2l}\psi_j}{dt^{2l}}\right),
\eea
which is derived from $\omega_M$ by taking into account the constraints (\ref{constr}).
The model (\ref{action}) is an $\,\mathcal{N}=1$ supersymmetric generalization of the free higher derivative particle studied in \cite{Gomis,Gonera_5,Gonera_4}.

In accord with (\ref{trN1}), the action (\ref{action}) is invariant under the transformations
\bea\label{tr}
\d t=a,\qquad \d x_i=\sum_{n=0}^{2l}\tilde{\zeta}_i^{(n)}t^n,\qquad \d\psi_i=\sum_{n=0}^{2l-1}\tilde{\xi}_i^{(n)}t^n,
\eea
where $\tilde{\zeta}_i^{(n)}=(-1)^n\zeta_i^{(n)}$, $\tilde{\xi}^{(n)}=(-1)^n\xi_i^{(n)}$. Then the Noether theorem yields the vector constants of the motion
\bea\label{vect}
C_i^{(n)}=\la_{ij}\sum_{k=0}^{n}\frac{(-1)^{k+1}n!}{(n-k)!}t^{n-k} x_j^{(2l-k)},\quad L_i^{(n)}=i\la_{ij}\sum_{k=0}^{n}\frac{(-1)^{k}n!}{(n-k)!}t^{n-k} \psi_j^{(2l-k-1)},
\eea
as well as $K_{-1}$
\bea\label{K1}
K_{-1}=\frac{1}{2}\la_{ij}\left(\sum_{k=1}^{2l}(-1)^{k+1}x_i^{(k)}x_j^{(2l-k+1)}+i\sum_{k=1}^{2l-1}(-1)^k \psi_i^{(k)}\psi_j^{(2l-k)}\right).
\eea
Here and in what follows the upper superscript in braces, which is attached to coordinates, denotes the number of time derivatives. Note that the latter is related to (\ref{vect}) via (\ref{biprod}).

Introducing the graded Poisson bracket
\bea\label{bracketG}
[A,B\}=\la_{ij}\sum_{n=0}^{2l}(-1)^n \frac{\partial A}{\partial x_i^{(n)}}\frac{\partial B}{\partial x_j^{(2l-n)}}+i\la_{ij}\sum_{n=0}^{2l-1}(-1)^{n+1}\frac{\overleftarrow{\partial}A}{\partial\psi_i^{(n)}}\frac{\overrightarrow{\partial}B}{\partial\psi_j^{(2l-n-1)}},
\eea
it is straightforward to check that the integrals of motion (\ref{biprod}) and (\ref{vect}) do obey the structure relations of the centrally extended $\mathcal{N}=1$ $l$-conformal Galilei superalgebra (\ref{algebraG}), (\ref{MCE}), (\ref{algebraGN1}), (\ref{MCEN1}) with $M=1$.
When verifying the algebra, the following relations
\bea\label{111}
[x_i^{(n)},x_j^{(m)}\}=(-1)^{n}\d_{n+m,2l}\la_{ij},\qquad [\psi_i^{(n)},\psi_j^{(m)}\}=i(-1)^{n+1}\d_{n+m,2l-1}\la_{ij}
\eea
prove to be helpful. It should be noted that the equations $[x_i^{(n)},K_{-1}\}=x_i^{(n+1)},\, [\psi_i^{(n)},K_{-1}\}=\psi_i^{(n+1)}$ hold.

Concluding this section, we display
the infinitesimal symmetry transformations of the action (\ref{action}) which correspond to the integrals of motion constructed above
\bea\label{transform}
&&
K_n:\qquad \d t=t^{n+1}a_n,\qquad \d x_i=l(n+1)t^n x_i a_n,\qquad \d\psi_i=(l-1/2)(n+1)t^n\psi_i a_n,
\nonumber\\[2pt]
&&
G_r:\qquad \d x_i=it^{r+1/2}\psi_i\alpha_r,\qquad\d\psi_i=(t^{r+1/2}\dot{x}_i-2l(r+1/2)x_i)\alpha_r,
\nonumber\\[2pt]
&&
M_{ij}:\quad\;\; \d x_i=w_{ij}x_j,\qquad\qquad\;\, \d\psi_i=w_{ij}\psi_j,\qquad (w_{ij}=-w_{ji}).
\eea

To summarize, a free superparticle obeying the higher derivative equations of motion (\ref{EOM}) provides the simplest dynamical realization of $\mathcal{N}=1$ $l$-conformal Galilei superalgebra. Note that this result is in agreement with the previous studies in \cite{Gomis}-\cite{Gonera_5} and \cite{Gonera_4}-\cite{Gonera_2}.

\vskip 0.5cm
\noindent
{\bf 4. Dynamical realization of $\mathcal{N}=1$ $l$-conformal Newton-Hooke superalgebra}
\vskip 0.5cm

As is known, realizations of the $l$-conformal Galilei algebra in a flat spacetime and in the Newton-Hooke spacetime \cite{Negro_1,Negro_2} are related by the coordinate transformations which, for the case of a negative cosmological constant, reads \cite{Duval_2,Galajinsky_3,Niederer}
\bea\label{nied}
t'=R\tan(t/R),\qquad x_i'(t')=x_i(t)/\cos^{2l}(t/R).
\eea
Here the prime designates coordinates parametrizing flat spacetime.

Let us construct an analogous transformation which links the model (\ref{action}) to its Newton-Hooke counterpart. To this end, we first note that the equations of motion for $\psi_i$ in (\ref{EOM}) can be formally obtained from the equations of motion for $x_i$ by the substitution $x_i\rightarrow\psi_i$, $l\rightarrow l-1/2$. Using this observation, one obtains the transformation for the odd variables
\bea\label{niedN1}
\psi_i'(t')=\psi_i(t)/\cos^{2l-1}(t/R).
\eea

Implementing the transformations (\ref{nied}) and (\ref{niedN1}) to (\ref{action}), one derives the action functional
\bea\label{PU1}
S=\frac{1}{2}\int dt\left(x_i\prod_{k=1}^{l+\frac{1}{2}}\left(\frac{d^2}{dt^2}+\frac{(2k-1)^2}{R^2}\right)x_i- i\psi_i\prod_{k=1}^{l-\frac{1}{2}}\left(\frac{d^2}{dt^2}+\frac{(2k)^2}{R^2}\right)\dot{\psi}_i\right),
\eea
which is valid for half-integer $l$. For integer $l$, one gets
\bea\label{PU2}
S=\frac{1}{2}\int dt\, \epsilon_{ij}\left(x_i\prod_{k=1}^{l}\left(\frac{d^2}{dt^2}+\frac{(2k)^2}{R^2}\right)\dot{x}_j- i\psi_i\prod_{k=1}^{l}\left(\frac{d^2}{dt^2}+\frac{(2k-1)^2}{R^2}\right)\psi_j\right).
\eea
These actions describe $\mathcal{N}=1$ supersymmetric extensions of the Pais-Uhlenbeck oscillator \cite{Pais} (for a review see \cite{Smilga}) for a particular choice of its frequencies. The bosonic limit of (\ref{PU1}) has been considered in detail in a recent work \cite{PU} (see also \cite{Galajinsky_1}).

The form of the symmetry transformations which leave the actions (\ref{PU1}) and (\ref{PU2}) invariant, as well as the form of the associated integrals of motion, is readily obtained by applying (\ref{nied}), (\ref{nied1}) to (\ref{tr}), (\ref{transform}), (\ref{vect}) and (\ref{biprod}), respectively. In particular, from (\ref{vect}) one derives the transformations corresponding to the vector generators
\bea\label{transf1}
\d x_i=\sum_{n=0}^{2l}\tilde{\zeta}_i^{(n)}R^n \sin^{n}{\frac{t}{R}}\cos^{2l-n}{\frac{t}{R}},\; \d\psi_i=\sum_{n=0}^{2l-1}\tilde{\xi}_i^{(n)}R^n \sin^{n}{\frac{t}{R}}\cos^{2l-n-1}{\frac{t}{R}},
\eea
and the integrals of motion associated with them
\bea\label{vect1}
C_i^{(n)}=\la_{ij}\sum_{k=0}^{n}\frac{(-1)^{k+1}n!}{(n-k)!}{(t')}^{n-k} ({x'}_j)^{(2l-k)},\; L_i^{(n)}=i\la_{ij}\sum_{k=0}^{n}\frac{(-1)^{k}n!}{(n-k)!}{(t')}^{n-k} ({\psi'}_j)^{(2l-k-1)}.
\eea
With respect to the graded bracket
\bea\label{bracketG1}
[A,B\}=\la_{ij}\sum_{n=0}^{2l}(-1)^n \frac{\partial A}{\partial ({x'}_i)^{(n)}}\frac{\partial B}{\partial ({x'}_j)^{(2l-n)}}+\la_{ij}\sum_{n=0}^{2l-1}(-1)^n\frac{\overleftarrow{\partial}A}{\partial({\psi'}_i)^{(n)}}\frac{\overrightarrow{\partial}B}{\partial({\psi'}_j)^{(2l-n-1)}},
\eea
they obey the structure relations (\ref{MCE}) and (\ref{MCEN1}). The existence of the relations similar to (\ref{biprod}) for the bosonic limit of (\ref{PU1}) was anticipated in \cite{Andrzejewski}. Eqs. (\ref{vect1}) and (\ref{bracketG1}) involve the derivatives of ${x'}_i$ and ${\psi'}_i$ with respect to $t'$, i.e., $\frac{d}{dt'}=\cos^{2}{\frac{t}{R}}\frac{d}{dt}$.

The remaining symmetries of the actions (\ref{PU1}) and (\ref{PU2}) read
\bea\label{transformPU}
&&
K_{0}:\qquad \d t=\frac{R}{2}\sin{\frac{2t}{R}}a_{0},\quad\;\,\d x_i=l\cos{\frac{2t}{R}}x_i a_{0},\quad\;\, \d\psi_i=\left(l-\frac{1}{2}\right)\cos{\frac{2t}{R}}\psi_i a_{0};
\nonumber
\\[2pt]
&&
K_{1}:\qquad \d t=R^2\sin^2{\frac{t}{R}}a_1,\quad \d x_i=lR\sin{\frac{2t}{R}}x_i a_1,\quad \d \psi_i=\left(l-\frac{1}{2}\right)R\sin{\frac{2t}{R}}\psi_i a_1;
\nonumber
\\[2pt]
&&
G_{-\frac{1}{2}}:\qquad \d x_i=i\cos{\frac{t}{R}}\psi_i\a_{-\frac{1}{2}},\qquad\d\psi_i=\left(\cos{\frac{t}{R}}\dot{x}_i +\frac{2l}{R}\sin{\frac{t}{R}}x_i\right)\a_{-\frac{1}{2}};
\nonumber
\\[2pt]
&&
G_{\frac{1}{2}}:\qquad\;\,\, \d x_i=iR\sin{\frac{t}{R}}\psi_i\a_{\frac{1}{2}},\qquad\d\psi_i=\left(R\sin{\frac{t}{R}}\dot{x}_i -2l\cos{\frac{t}{R}}x_i\right)\a_{\frac{1}{2}},
\eea
while the transformations corresponding to the generators $K_{-1}$ and $M_{ij}$  are the same as in (\ref{transform}). In order to obtain the transformation corresponding to $K_{-1}$ and the corresponding integral of motion, one has to make the linear change of the basis (\ref{change1}).
Thus, the actions (\ref{PU1}) and (\ref{PU2}) are invariant under $\mathcal{N}=1$ $l$-conformal Galilei superalgebra realized in Newton-Hooke superspace.

The case of a positive cosmological constant can be treated by implementing the formal change of the characteristic time $R\rightarrow iR$. In particular, for a half-integer $l$ the analogues of (\ref{nied}) and (\ref{niedN1}),
\bea\label{nied1}
t'=R\tanh(t/R),\qquad x_i'(t')=x_i(t)/\cosh^{2l}(t/R),\qquad \psi_i'(t')=\psi_i(t)/\cosh^{2l-1}(t/R),
\eea
yield the action functional
\bea\label{PU11}
S=\frac{1}{2}\int dt\left(x_i\prod_{k=1}^{l+\frac{1}{2}}\left(\frac{d^2}{dt^2}-\frac{(2k-1)^2}{R^2}\right)x_i- i\psi_i\prod_{k=1}^{l-\frac{1}{2}}\left(\frac{d^2}{dt^2}-\frac{(2k)^2}{R^2}\right)\dot{\psi}_i\right),
\eea
while for integer $l$ one gets
\bea\label{PU21}
S=\frac{1}{2}\int dt\, \epsilon_{ij}\left(x_i\prod_{k=1}^{l}\left(\frac{d^2}{dt^2}-\frac{(2k)^2}{R^2}\right)\dot{x}_j- i\psi_i\prod_{k=1}^{l}\left(\frac{d^2}{dt^2}-\frac{(2k-1)^2}{R^2}\right)\psi_j\right).
\eea
These actions can be viewed as describing $\mathcal{N}=1$ supersymmetric extensions of the Pais-Uhlenbeck oscillator for the particular choice of imaginary frequencies.

It should be stressed that in the flat space limit $R\rightarrow \infty$ all the formulas corresponding to the realizations (\ref{PU1}), (\ref{PU2}), (\ref{PU11}), (\ref{PU21}) in the Newton-Hooke superspace correctly reproduce the respective expressions in a flat superspace.

\vskip 0.5cm
\noindent
{\bf 5. Conclusion}
\vskip 0.5cm
To summarize, in this work we have constructed dynamical realizations of $\mathcal{N}=1$ $l$-conformal Galilei superalgebra in a flat superspace and in the Newton-Hooke superspace. Coordinate transformations which link the realizations were found. The models describe a free higher derivative superparticle and an $\mathcal{N}=1$ supersymmetric extension of the Pais-Uhlenbeck oscillator for the particular choice of its frequencies, respectively.

Turning to possible further developments, it is interesting to investigate whether an $\mathcal{N}=1$ supersymmetric extension of the $l$-conformal Galilei algebra can be realized in systems without higher derivatives in the equations of motion. In this context, it would be interesting to construct a supersymmetric extensions of the models constructed in \cite{Galajinsky_2}, \cite{Galajinsky_1}. Then it is worth generalizing the results in \cite{Henkel_2}, \cite{Henkel_3} to incorporate $\mathcal{N}=1$ supersymmetry. A generalization of the analysis in this paper to the case of $\mathcal{N}=2$ $l$-conformal Galilei superalgebras \cite{Masterov,Aizawa_1,Aizawa_2} is worth studying as well.

\vskip 0.5cm
\noindent
{\bf Acknowledgements}
\vskip 0.5cm
We thank A. Galajinsky for the comments on the paper. This work was supported by the Dynasty Foundation, Russian Foundation for Basic Research (RFBR) grant 14-02-31139-Mol, the MSE program "Nauka" under project 3.825.2014/K, and TPU grant LRU.FTI.123.2014.

\vskip 0.5cm
\noindent
{\bf Appendix A. Infinite-dimensional extension.}
\vskip 0.5cm

It should be noted that the generators $K_n$ and $G_r$ in (\ref{algebraGN1}) form centerless $\mathcal{N}=1$ Neveu-–Schwarz algebra ($\mathcal{NS}$) \cite{NS}, if one takes $n$ and $r$ to be arbitrary integer and half-integer numbers, respectively. It is natural to ask whether an infinite-dimensional extension of the $\mathcal{N}=1$ $l$-conformal Galilei superalgebra is feasible. Such an extension for $\mathcal{N}=1$ Schr\"{o}dinger superalgebra in $d=1$ has been systematically constructed in \cite{Henkel_1} as the superalgebra of some functions forming a closed set under the graded Poisson bracket

$$
[f,g\}=\frac{\partial f}{\partial q}\frac{\partial g}{\partial p}-\frac{\partial f}{\partial p}\frac{\partial g}{\partial q}+i \frac{\overleftarrow{\partial}f}{\partial\theta}\frac{\overrightarrow{\partial}g}{\partial\theta}.\eqno{(A1)}
$$

According to the analysis in \cite{Henkel_1}, one can choose $K_{n}=q\,p^{n+1}$ and $G_r=\sqrt{2}\,q^{\frac{1}{2}}\,p^{r+\frac{1}{2}}\,\theta$ which under (A1) obey the structure relations of $\mathcal{NS}$. Other generators
$$
C^{(n)}=q^{l}\,p^{n},\qquad L^{(n)}=\frac{1}{\sqrt{2}}\,q^{l-1/2}\,p^{n}\,\theta.\eqno{(A2)}
\nonumber
$$
can be unambiguously fixed by imposing the structure relations (\ref{algebraGN1}). In contrast to (\ref{MCE}) and (\ref{MCEN1}) the nonvanishing relations between the generators (A2) read
$$
[C^{(n)},C^{(m)}\}=l(m-n)M^{m+n}_{1,1},\qquad[L^{(n)},L^{(m)}\}=\frac{i}{2}M^{n+m+1}_{1,1},\eqno{(A3)}
$$
where we denoted
$$
M^{n}_{k,s}=p^{n-1}q^{2lk-s}.\eqno{(A4)}
$$
Moreover, the relations involving (A4) and $G_{r}$, $L^{(n)}$,
$$
[G_r,M^{n}_{k,s}\}=(n-1-(2lk-s)(2r+1))\tilde{M}^{n+r-1}_{k,2s+1},
$$
$$
[L^{(n)},M^{m}_{k,s}\}=\frac{1}{2}\left((2l-1)(m-1)-2n(2lk-s)\right)\tilde{M}^{m+n-3/2}_{k,2s-2l+3},\eqno{(A5)}
$$
yield extra generators
$$
\tilde{M}^{r}_{k,s}=\frac{1}{\sqrt{2}}p^{r-\frac{1}{2}}q^{2lk-\frac{1}{2}s}\theta.\eqno{(A6)}
$$
Along with (\ref{algebraG}), (\ref{algebraGN1}), (A3), (A5) the structure relations of the desired infinite-dimensional extension involve
\bea
\begin{aligned}
&
[M^{n_1}_{k_1,s_1},M^{n_2}_{k_2,s_2}\}=\left((2lk_1-s_1)(n_2-1)-(2lk_2-s_2)(n_1-1)\right)M^{n_1+n_2-2}_{k_1+k_2,s_1+s_2+1},\;
\\[2pt]
&
[K_{n},M^{m}_{k,s}\}=(m-1-(2lk-s)(n+1))M^{n+m}_{k,s},\quad\;\; [\tilde{M}_{k_1,s_1}^{r_1},\tilde{M}_{k_2,s_2}^{r_2}\}=\frac{i}{2}M_{k_1+k_2,\frac{1}{2}(s_1+s_2)}^{r_1+r_2},
\\[2pt]
&
[K_n,\tilde{M}^{r}_{k,s}\}=(r-1/2-(2lk-s/2)(n+1))\tilde{M}^{n+r}_{k,s},\;[G_{r_1},\tilde{M}^{r_2}_{k,s}\}=iM^{r_1+r_2}_{k,\frac{1}{2}(s-1)},
\\[2pt]
&
[C^{(n)},M^{m}_{k,s}\}=\left(l(m-1)-n(2lk-s)\right)M^{n+m-1}_{k,s-l+1},\quad\quad [L^{(n)},\tilde{M}^{r}_{k,s}\}=\frac{i}{2}M^{n+r+1/2}_{k,\frac{1}{2}(s+1)-l},
\\[2pt]
&
[C^{(n)},\tilde{M}^{r}_{k,s}\}=\left(l(r-1/2)-n(2lk-s/2)\right)\tilde{M}^{n+r-1}_{k,s-2l+2},
\\[2pt]
&
[L^{(n)},M^{m}_{k,s}\}=\frac{1}{2}\left((2l-1)(m-1)-2n(2lk-s)\right)\tilde{M}^{m+n-3/2}_{k,2s-2l+3}.
\nonumber
\end{aligned}
\eea

By analogy with \cite{Henkel_1}, one can define the grading of coordinates as follows
$$
gra(p)=0,\qquad gra(q)=1,\qquad gra(\theta)=\frac{1}{2},
$$
with respect to which
$$
gra(K_n)=gra(G_r)=1,\qquad gra(C^{(n)})=gra(L^{(n)})=l,
$$
$$
gra(M^{n}_{k,s})=2lk-s,\quad gra(\tilde{M}^{r}_{k,s})=2lk-s/2+1/2.\eqno{(A7)}
$$
In Ref. \cite{Henkel_1}, the grading relations (A7) play the crucial role as they allow one to simplify the infinite-dimensional superalgebra considerably. In particular, it has been shown that for $l=1/2$ one can construct the quotient algebra which involves only additional generators $M^{n}_{1,1}$ and $\tilde{M}^{n}_{1,3}$ (for more details see \cite{Henkel_1}). Unfortunately, the difference in grading (A7) makes an analogous simplification for the superalgebra with $l>1$  problematic.

\end{document}